\documentclass[12pt,a4paper,english,aps,pre,preprint,floatfix]{revtex4}
\usepackage[T1]{fontenc}
\usepackage[latin1]{inputenc}
\usepackage{graphicx}

\makeatletter

\usepackage{babel}
\makeatother
\begin{document}

\title{Sivashinsky equation in a rectangular domain}

\author{Bruno Denet}

\affiliation{IRPHE 49 rue Joliot Curie BP 146 Technopole de Chateau Gombert 13384
Marseille Cedex 13 France}

\email{bruno.denet@irphe.univ-mrs.fr}

\pacs{47.54.+r 47.70.Fw 47.20.Ky}

\preprint{submitted to Phys. Rev. E}

\begin{abstract}
The (Michelson) Sivashinsky equation of premixed flames is studied
in a rectangular domain in two dimensions. A huge number of 2D stationary
solutions are trivially obtained by addition of two 1D solutions.
With Neumann boundary conditions, it is shown numerically that adding
two stable 1D solutions leads to a 2D stable solution. This type of
solution is shown to play an important role in the dynamics of the
equation with additive noise.
\end{abstract}
\maketitle

\section{Introduction \label{sec:Introduction}}

The Sivashinsky equation \cite{siva77} (or Michelson Sivashinsky
equation depending on the authors) is a non linear equation which
describes the time evolution of premixed flames. Because of the jump
of temperature (and thus of density) across the flame, a plane flame
front is submitted to a hydrodynamic instability called the Darrieus-Landau
instability. The conservation of normal mass flux and tangential velocity
across the front leads to a deflection of streamlines which is the
main cause of this instability. A more detailed description of this
instability can be found in the book of Williams \cite{williamsbook}
(see also, in the approximation of potential flow in the burnt gases,
the elementary electrostatic explanation in \cite{denetbunsen2d},
where essentially the flame is described as a surface with a uniform
charge density). At small scales, the instability is damped by diffusive
effects: the local front propagation velocity is modified by a term
proportional to curvature, the coefficient ahead of the curvature
term is called the Markstein length. A geometrical non linear term,
which limits ultimately the growth of the instability, is caused by
the normal propagation of the flame. The Sivashinsky equation, obtained
as a development in powers of a gas expansion parameter, i.e. for
a small jump of temperature, or equivalently for a flow almost potential
in the burnt gases, represents a balance between the evolution due
to these three effects, Darrieus Landau instability, stabilization
by curvature and normal propagation of the flame.

The qualitative agreement between the Sivashinsky equation and direct
numerical simulations, generally performed with periodic boundary
conditions has been excellent even with large gas expansion \cite{rastigejevmatalonjfm},
and also when gravity is included \cite{denetbonino}. It has been
shown in a classic paper of the field \cite{thualfrischhenon} (following
\cite{leechen}, where the pole decomposition was introduced) that
the 1D solution of the Sivashinsky equation in the absence of noise
was attracted for large times toward stationary solutions, with poles
aligned in the complex plane, called coalescent solutions. It was
shown analytically in \cite{vaynblattmatalon1}\cite{vaynblattmatalon2}
that each solution, with a given number of poles, is linearly stable
in a given interval for the control parameter (either the domain width
or more often the Markstein length with a domain width fixed to $2\pi$).

In a recent paper \cite{denetstationary} (hereafter called I) the
present author has been interested in the behavior of the Sivashinsky
equation in 1D, but with Neumann boundary conditions (zero slope at
each end of the domain), a situation which, although more realistic
than periodic boundary conditions, had not attracted much interest
over the years. Actually, periodic boundary conditions lead to a symmetry
which is not present in the case of a flame in a tube, i.e. every
lateral translation of a given solution is also a solution. Presented
in a different way, a perturbation on the flame can grow, reach the
cusp (the very curved part of the front, pointing toward the burnt
gases) and then decay, but after having caused a global translation
of the original solution. This is not possible with Neumann boundary
conditions, but it was supposed that this difference with periodic
boundary conditions was unimportant. The surprise was however that
stable stationary solutions in the Neumann case involved a number
of solutions with two cusps (and the corresponding poles) at each
end of the domain, called bicoalescent solutions. This type of solution
of the Sivashinsky equation was already introduced in \cite{guidimarchetti},
although this last article did not obtain those which are stable with
Neumann boundary conditions. The author would like to mention here
two articles which he did not cite in I , namely \cite{gutmansiva},
where some bicoalescent solutions with Neumann boundary conditions
were first obtained, and \cite{Travnikov-et.al-00} where bicoalescent
solutions were obtained in direct numerical simulations. In this last
paper, one solution was not completely stationary, because of the
effect of noise, but another solution was actually almost stationary.
Of course the computer time needed for such a simulation is probably
one hundred times more than the equivalent Sivashinsky equation simulation,
with all sorts of possible sources of noise, so obtaining really stationary
bicoalescent solutions in this case is a challenging task.

Coming back to I, we can summarize the 1D results of this paper in
the following way:

\begin{enumerate}
\item Bicoalescent solutions were obtained, stable with Neumann boundary
conditions. Simulations performed without noise tend to these solutions.
\item The new solutions led to a bifurcation diagram with a large number
of stationary solutions, where particularly the number of solutions
multiply when the Markstein length, presented above, which controls
the stabilizing influence of the curvature term, decreases
\item The bicoalescent solutions play a major role in the dynamics of the
equation with additive noise. In the case of moderate white noise,
the dynamics is controlled by jumps between different bicoalescent
solutions.
\end{enumerate}
In the present paper, we shall be interested in the Sivashinsky equation
with Neumann boundary conditions, but in two dimensions in a rectangular
domain. Another nice property of the equation (apart from the pole
decomposition in 1D) is that 2D solutions can be formed by the simple
addition of two 1D solutions, one for each coordinate \cite{joulinimagemaths}.
The exact counterpart of I will be obtained:

\begin{enumerate}
\item Sums of two bicoalescent solutions are stable in 2D with Neumann boundary
conditions. The time evolution of the equation without noise tends
toward these solutions
\item With sums of a large number of 1D solutions, a really huge number
of 2D solutions can be obtained.
\item The sums of bicoalescent solutions play also a major role in the dynamics
in two dimensions in the presence of noise.
\end{enumerate}

\section{Solutions in elongated domains \label{sec:elongated}}

The Sivashinsky equation in one dimension can be written as 

\begin{equation}
\phi_{t}+\frac{1}{2}\phi_{x}^{2}=\nu\phi_{xx}+I\left(\phi\right)\label{eq:sivashinsky}\end{equation}

where $\phi\left(x\right)$ is the vertical position of the front.
The Landau operator $I\left(\phi\right)$ corresponds to a multiplication
by $\left|k\right|$ in Fourier space, where $k$ is the wavevector,
and physically to the destabilizing influence of gas expansion on
the flame front (known as the Darrieus-Landau instability, and described
in the introduction). $\nu$ is the only parameter of the equation
(the Markstein length) and controls the stabilizing influence of curvature.
The linear dispersion relation giving the growth rate $\sigma$ versus
the wavevector is, including the two effects: 

\begin{equation}
\sigma=\left|k\right|-\nu k^{2}\label{eq:dispersion}\end{equation}

As usual with Sivashinsky-type equations, the only non linear term
added to the equation is $\frac{1}{2}\phi_{x}^{2}$. In the flame
front case, this term is purely geometrical : the flame propagates
in the direction of its normal, a projection on the vertical ($y$)
direction gives the factor $\cos\left(\theta\right)=1/\sqrt{1+\phi_{x}^{2}}$,
where $\theta$ is the angle between the normal and the vertical direction,
then a development valid for small slopes of the front leads to the
term $\frac{1}{2}\phi_{x}^{2}$. The Sivashinsky equation is typically
solved numerically on $[0,2\pi]$ with periodic boundary conditions.
In I it has also been solved on $[0,2\pi]$ with only symmetric modes,
which corresponds to homogeneous Neumann boundary conditions on $[0,\pi]$
(zero slope on both ends of the domain). The two dimensional version
of the Sivashinsky equation is 

\begin{equation}
\phi_{t}+\frac{1}{2}\left(\nabla\phi\right)^{2}=\nu\Delta\phi+I\left(\phi\right)\label{eq:sivashinsky2d}\end{equation}

where the Landau operator $I\left(\phi\right)$ corresponds now to
a multiplication by $\sqrt{k_{x}^{2}+k_{y}^{2}}$ in Fourier space,
$k_{x}$ and $k_{y}$ being the wavevectors in the $x$ and $y$ directions.
All dynamical calculations, are performed by Fourier pseudo-spectral
methods (i.e. the non linear term is calculated in physical space
and not by a convolution product in Fourier space). The method used
is first order in time and semi-implicit (implicit on the linear terms
of the equation, explicit on $\frac{1}{2}\phi_{x}^{2}$). No particular
treatment of aliasing errors is used. The 2D Sivashinsky equation
is solved in $[0,2\pi]*[0,2b]$ with only symmetric modes, which corresponds
to homogeneous Neumann boundary conditions in the rectangular domain
$[0,\pi]*[0,b]$ .

Pole solutions (\cite{thualfrischhenon}) of the 1D Sivashinsky equation
are solutions of the form:

\begin{equation}
\phi=2\nu\sum_{n=1}^{N}\left\{ \ln\left(\sin\left(\frac{x-z_{n}(t)}{2}\right)\right)+\ln\left(\sin\left(\frac{x-z_{n}^{*}(t)}{2}\right)\right)\right\} \label{eq:poledecomposition}\end{equation}

where $N$ is the number of poles $z_{n}(t)$ in the complex plane.
Actually the poles appear in complex conjugate pairs, and the asterisk
in Equation \ref{eq:poledecomposition} denotes the complex conjugate.
In all the paper, the number of poles will also mean number of poles
with a positive imaginary part. The pole decomposition transforms
the solution of the Sivashinsky equation into the solution of a dynamical
system for the locations of the poles. In the case of stationary solutions,
the locations of the poles are obtained by solving a non linear system:

\begin{equation}
-\nu\sum_{l=1,l\ne n}^{2N}\cot\left(\frac{z_{n}-z_{l}}{2}\right)-i\textnormal{sgn}\left[{\textstyle \textnormal{Im}}\left(z_{n}\right)\right]=0\;\;\;\;\; n=1,\cdots,N\label{eq:nonlinearsystem}\end{equation}

where $\textnormal{Im}\left(z_{n}\right)$ denotes the imaginary part
and sgn is the signum function. This non linear system is solved by
a variant of Newton method.

With periodic boundary condition, the usual result is that in the
window $2n-1$$\leq1/\nu$$\leq2n+1$, $n=1,2,\cdots$ there exists
$n$ different monocoalescent stationary solutions (all the poles
have the same real part), with $1$ to $n$ poles, and the solution
with the maximum number of poles $n$ is asymptotically stable. For
a particular value of $1/\nu$, the number $n(\nu)$ such that $2n-1$$\leq1/\nu$$\leq2n+1$
is called the optimal number of poles. 

With Neumann boundary conditions, in each of the intervals $\left[2n-1,2n+1\right]$
of the parameter $1/\nu$, not only one asymptotically stable solution,
but $n+1$, of the form $\left(l,n-l\right)$ with $l=0,1,$$\cdots,n$
where $l$ poles coalesce at $x=0$ and $l-n$ coalesce at $x=\pi$,
were obtained in I. (The bicoalescent type of solutions have been
recently introduced in \cite{guidimarchetti}). In Figure \ref{fig:stable_1d}
is shown a bifurcation diagram with all the possible stable stationary
solutions (plotted only, contrary to I, in the domain where they are
stable) versus $1/\nu$. What is actually plotted is the amplitude
$\Delta\phi$ (maximum minus minimum of $\phi$) versus $1/\nu.$
As can be seen, when the optimal number of poles increases with $1/\nu$,
the number of stable stationary bicoalescent solutions is also increasing.
The stability of these solutions is not proved analytically, nor by
a numerical study of the linearized problem, we use only numerical
simulations of the Sivashinsky equation, with the different bicoalescent
solutions plus some small perturbations as initial conditions, and
the solution returns toward the unperturbed solution.

In a square domain $[0,2\pi]*[0,2\pi]$, it has been remarked in \cite{joulinimagemaths}
that if $\phi_{1}(x)$ and $\phi_{2}(x)$ are solutions of the 1D
Sivashinsky equation (\ref{eq:sivashinsky}), then $\phi_{1}(x)+\phi_{2}(y)$
(we use here $\phi_{1}\oplus\phi_{2}$ as a notation for this sum,
whose amplitude is the sum of the amplitudes of $\phi_{1}$ and $\phi_{2}$)
is a solution of (\ref{eq:sivashinsky2d}) in two dimensions, and
that the stationary solution obtained numerically in this case for
periodic boundary conditions \cite{michelsonsiva} is simply a sum
of two monocoalescent 1D solutions. Let us note that, if it is absolutely
obvious that sums are solutions of the 2D equation, the stability
of these solutions has never been proved analytically, and can only
be inferred from a small number of numerical simulations.

In the case of rectangular domains $[0,2\pi]*[0,2b]$, sums are also
solutions of the equation, with $\phi_{2}$ now solution of the 1D
Sivashinsky equation with parameter $1/\nu$ in domain $[0,2b]$,
which can be obtained by an appropriate rescaling from the solution
in $[0,2\pi]$ with parameter $1/\nu_{1}=(1/\nu)(b/\pi)$. 

\begin{figure}[t]

\caption{\label{fig:3ribbons}Perspective view of the (from top to bottom)
$\left(5,0\right)\oplus(0)$, $\left(4,1\right)\oplus(0)$, $\left(3,2\right)\oplus(0)$
stationary solutions for $1/\nu=10$ and $b=\pi/10$. The solution
is plotted in the interval $[0,2\pi]*[0,2b]$ because it is easier
to visualize. Actually, Neumann boundary conditions are satisfied
in $[0,\pi]*[0,b]$ (one fourth of the domain shown).}

\includegraphics[%
  scale=0.8]{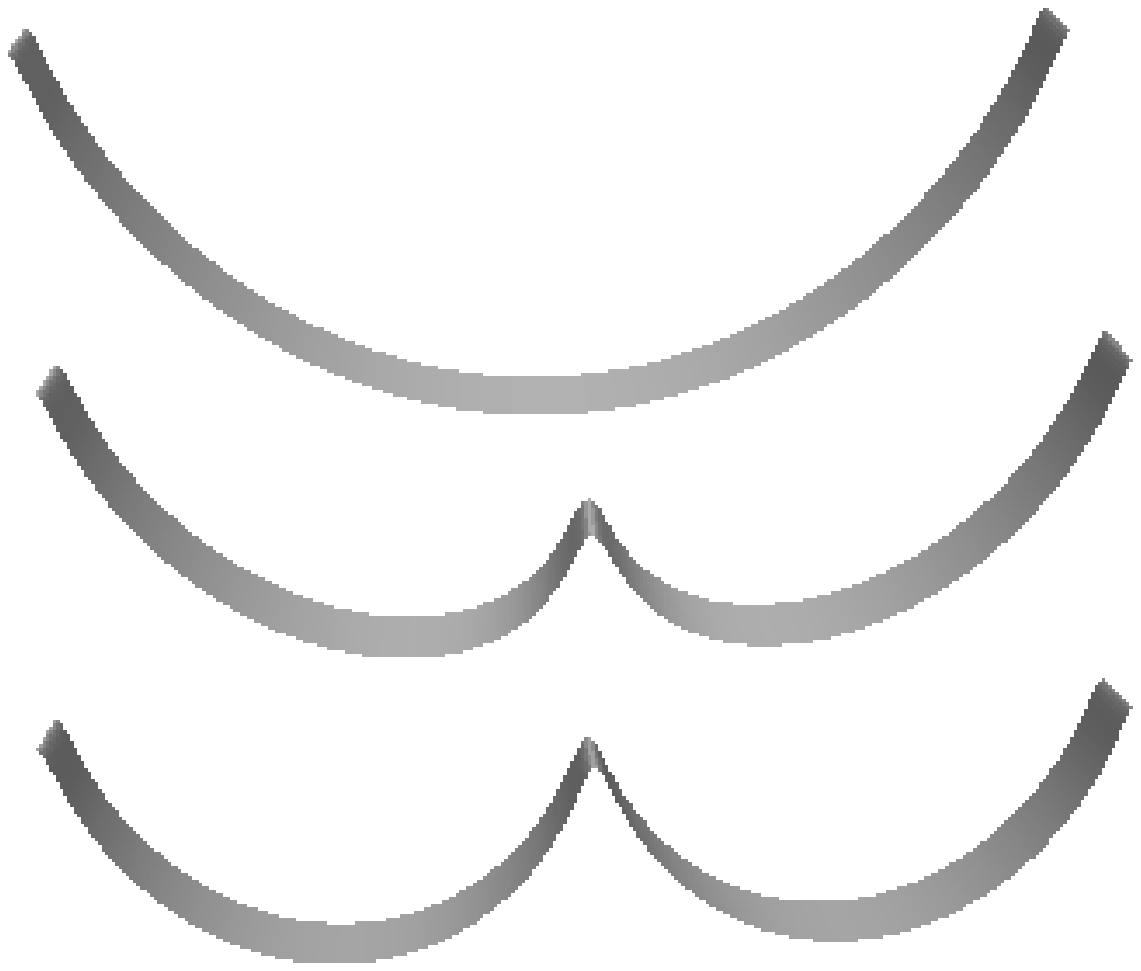}
\end{figure}

A particularly simple case is the limit where $b$ is very small,
where the only solution with parameter $1/\nu$ in domain $[0,2b]$
is simply the flat $(0)$ solution $\phi_{2}=0$. As a sum of the
previously described bicoalescent solutions in $[0,2\pi]$ added to
the flat solution in the other direction, we have simply a way to
observe the bicoalescent solutions in two dimensions. We have observed
numerically (not shown here, the behavior is very similar to the 1D
case) for Neumann boundary conditions, that these sums $\left(l,n-l\right)\oplus(0)$
are stable . As an example, for $1/\nu=10$ and $b=\pi/10$ we show
in Figure \ref{fig:3ribbons} a perspective view of the three different
stationary bicoalescent solutions $\left(5,0\right)\oplus(0)$, $\left(4,1\right)\oplus(0)$,
$\left(3,2\right)\oplus(0)$ (from top to bottom). In all the figures,
the whole domain $[0,2\pi]*[0,2b]$ is plotted, the solution with
Neumann boundary conditions corresponds only to one fourth of the
domain $[0,\pi]*[0,b]$ . We have found it clearer to show the whole
domain (contrary to I), because some solutions are very difficult
to distinguish if plotted in $[0,\pi]*[0,b]$. Although these solutions
are very sensitive to noise (although less than the pure 1D solutions)
it could be possible to observe in direct numerical simulations and
experimentally the solutions with the lower amplitude, which are the
least sensitive to noise. In experiments, the solutions should also
survive heat losses (important in narrow channels) and not be too
much perturbed by gravity (i.e. have a large enough Froude number)
in order to be observed .

\section{solutions in square domains \label{sec:square}}

\begin{figure}[p]

\caption{\label{fig:stable_1d} Stable stationary solutions in 1D: amplitude
$\Delta\phi$ vs $1/\nu$. All the different branches are only plotted
for the values of $1/\nu$ where they are stable. A notation like
(3,2) means that 3 poles are located at $x=0$, and 2 poles at $x=\pi$}

\includegraphics[%
  bb=4bp 153bp 694bp 613bp,
  scale=0.6]{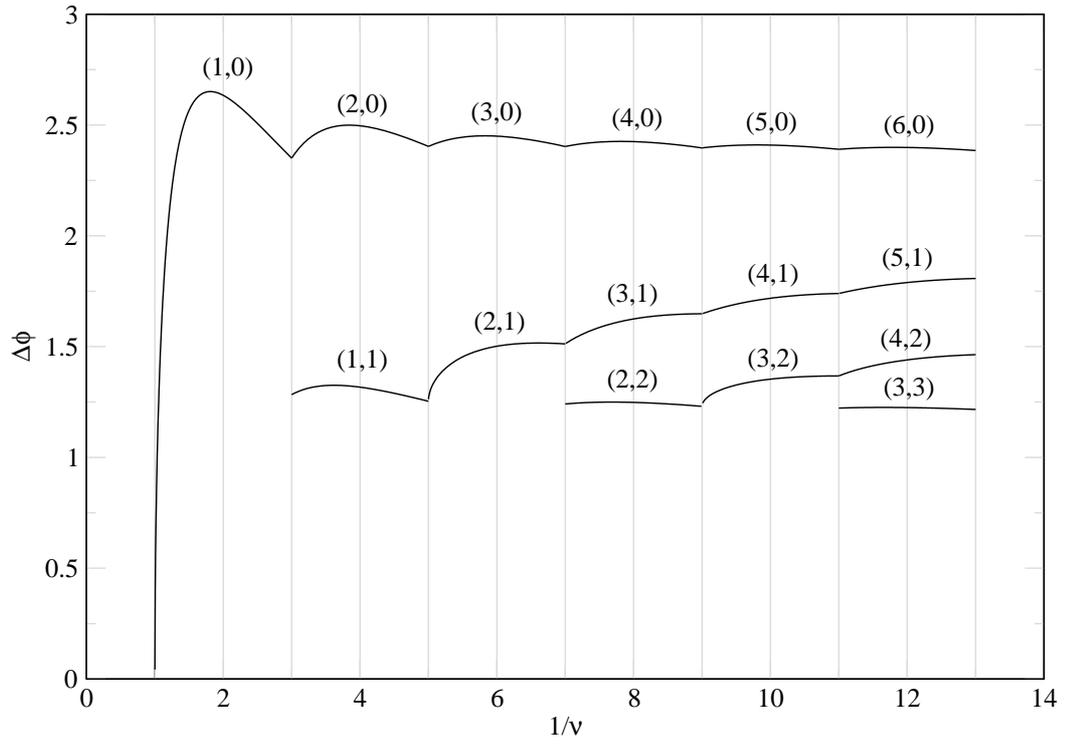}
\end{figure}

We now turn to stationary solutions in square domains $[0,2\pi]*[0,2\pi]$
with Neumann boundary conditions. Sums of bicoalescent solutions produce
also in this case stable stationary solutions. The purpose of this
section is to give details on the consequences of this simple addition
property. We show first the different types of solutions obtained
by addition of stable bicoalescent solutions in 1D. We insist on the
fact that these solutions are linearly stable and give a specific
example of the time evolution of one such solution with some small
perturbations. Finally two bifurcation diagrams are provided, one
is the 2D equivalent of Figure \ref{fig:stable_1d} with the stable
solutions plotted only in their stable domain. The second contains
all the solutions obtained by addition of all the branches found in
1D in I, and as the reader will see, a really huge number of branches
are created in this way.

\begin{figure}

\caption{\label{fig:3_others}Perspective view of the (from top to bottom)
$\left(4,1\right)\oplus(4,1)$, $\left(4,1\right)\oplus(3,2)$, $\left(3,2\right)\oplus(3,2)$
stationary solutions for $1/\nu=10$. The solution is plotted in the
interval $[0,2\pi]*[0,2\pi]$ because it is easier to visualize. Actually,
Neumann boundary conditions are satisfied in $[0,\pi]*[0,\pi]$ (one
fourth of the domain shown).}

\includegraphics[%
  bb=41bp 53bp 686bp 613bp,
  scale=0.7]{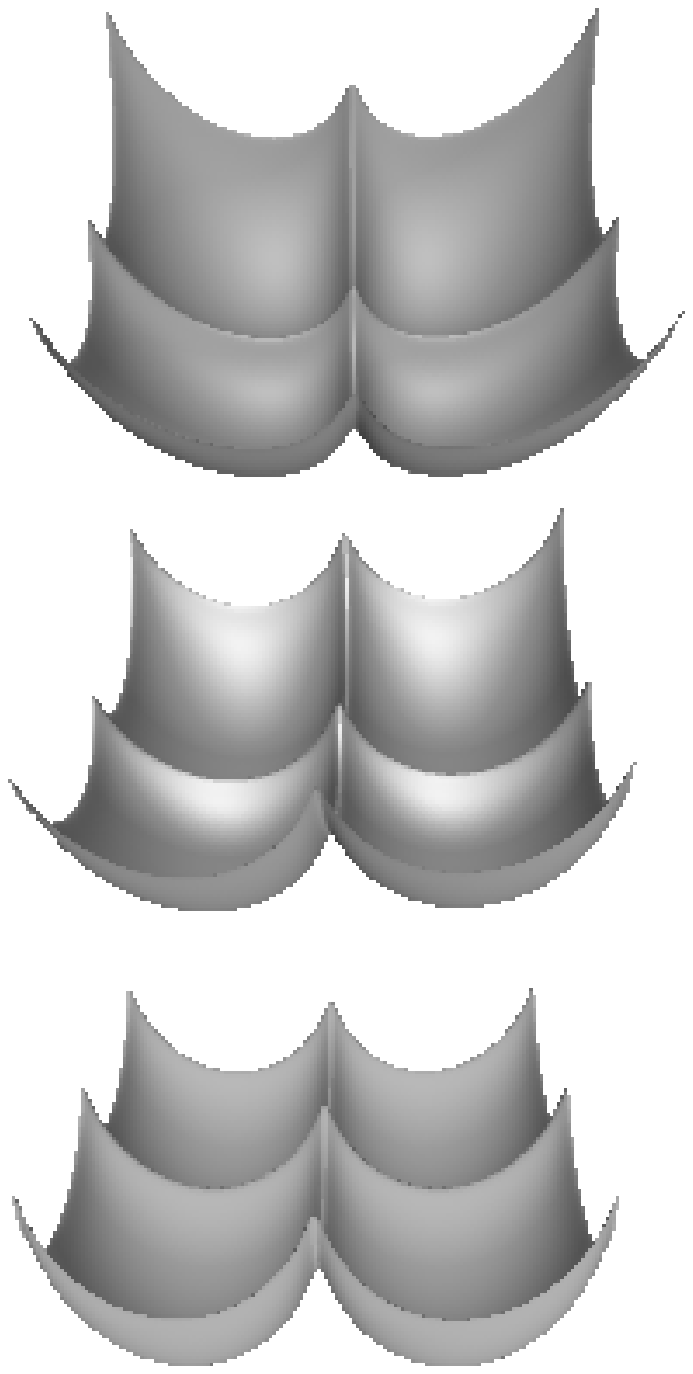}
\end{figure}

In Figures \ref{fig:3_5,0} and \ref{fig:3_others} are shown the
six stable solutions obtained from the three 1D solutions of Figure
\ref{fig:3ribbons} for $1/\nu=10$. In Figure \ref{fig:3_5,0} can
be seen (in perspective view, for the whole domain $[0,2\pi]*[0,2\pi]$),
from top to bottom the $\left(5,0\right)\oplus(5,0)$, $\left(4,1\right)\oplus(5,0)$and
$\left(3,2\right)\oplus(5,0)$ solutions. In Figure \ref{fig:3_others}
can be seen the three remaining solutions $\left(4,1\right)\oplus(4,1)$,
$\left(3,2\right)\oplus(4,1)$ and $\left(3,2\right)\oplus(3,2)$.
All these solutions are found to be linearly stable, although all
the solutions of Figure \ref{fig:3_5,0} ($(5,0)\oplus$ something)
are extremely sensitive to noise. It must be pointed out that most
of these solutions would have been almost impossible to find from
a time integration of the 2D Sivashinsky equation (Equation (\ref{eq:sivashinsky2d}))
because of this sensitivity to noise, and it is likely that obtaining
them from a steady version of (\ref{eq:sivashinsky2d}) would have
been very difficult too.

In Figure \ref{fig:ampli_3,2_4,1}, we have an example showing the
stability of the $\left(3,2\right)\oplus(4,1)$ solution. We start
from this solution and add an additive white noise to Equation (\ref{eq:sivashinsky2d})
when the time is below $0.5$. This white noise is gaussian, of deviation
one, and we multiply it by an amplitude $a=0.001$. It can be seen
that after the noise is stopped, the solution tends exponentially
back toward the $\left(3,2\right)\oplus(4,1)$ solution. Similar figures
would be obtained with the other solutions of Figures \ref{fig:3_5,0}
and \ref{fig:3_others}, except that higher amplitude solutions would
need an even lower noise in order not to jump immediately toward a
lower amplitude solution.

\begin{figure}

\caption{\label{fig:stable_2d}Stable stationary solutions in 2D for a square
domain : amplitude $\Delta\phi$ vs $1/\nu$. All the different branches
are only plotted for the values of $1/\nu$ where they are stable.
The 2D linearly stable solutions are obtained by addition of the corresponding
1D linearly stable solutions of Figure \ref{fig:stable_1d}.}

\includegraphics[%
  bb=57bp 70bp 508bp 608bp,
  scale=0.6]{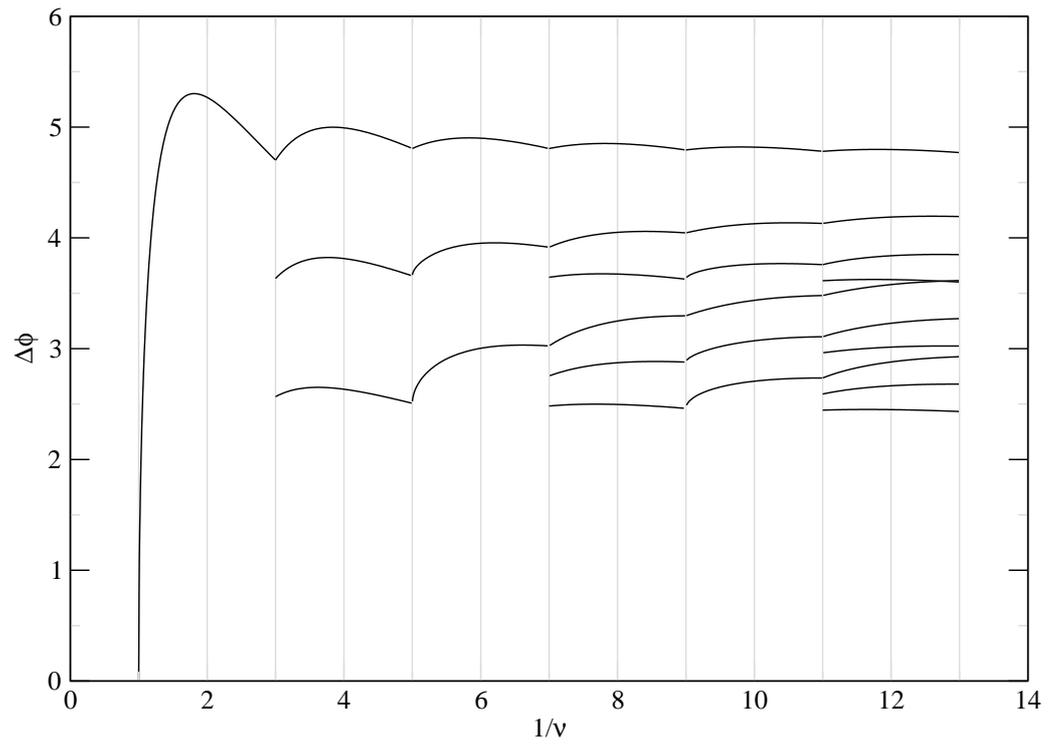}
\end{figure}

In Figure \ref{fig:stable_2d} is shown the strict 2D equivalent of
Figure \ref{fig:stable_1d}: the bifurcation diagram showing the amplitude
versus $1/\nu$ for all the solutions which are linearly stable, only
plotted in their domain of stability. For $1/\nu<3$ there is only
one possibility $\left(1,0\right)\oplus(1,0)$. For $3<1/\nu<5$ we
have three branches (from higher to lower amplitudes) $\left(2,0\right)\oplus(2,0)$
$\left(1,1\right)\oplus(2,0)$ and $\left(1,1\right)\oplus(1,1)$.
For the value $1/\nu=10$ we have the six solutions of Figures \ref{fig:3_5,0}
and \ref{fig:3_others}, that is from higher to lower amplitudes the
$\left(5,0\right)\oplus(5,0)$, $\left(4,1\right)\oplus(5,0)$ $\left(3,2\right)\oplus(5,0)$
$\left(4,1\right)\oplus(4,1)$, $\left(3,2\right)\oplus(4,1)$ and
$\left(3,2\right)\oplus(3,2)$ solutions. Higher values of $1/\nu$
would correspond to an increasing number of stable stationary solutions.

Naturally, neither Figure \ref{fig:stable_1d} (in 1D) or Figure \ref{fig:stable_2d}
(in 2D) contain all the possible stationary solutions. In 1D Guidi
and Marchetti \cite{guidimarchetti} have introduced the concept of
interpolating solutions, which are unstable solutions connecting different
branches of stable solutions in the previous bifurcation diagrams.
In I, the present author has shown that this leads to a complex network
of solutions, which was called web of stationary solutions. But now
in two dimensions, we have the possibility, when two branches $\phi_{1}$
and $\phi_{2}$ exist for a parameter $1/\nu$ to create the 2D branch
$\phi_{1}\oplus\phi_{2}$. This construction leads to a bifurcation
diagram (with as before $1/\nu<14$, i.e. not very large flames) with
a truly huge number of different stationary solutions (several thousands
of branches). The comparison with Figure \ref{fig:stable_2d} shows
that most of these solutions are linearly unstable.

\begin{figure}

\caption{\label{fig:all_2d}Stationary solutions in 2D for a square domain
: amplitude $\Delta\phi$ vs $1/\nu$ (figure with all the solutions
obtained by addition of 1D stationary solutions). When two 1D branches
found in I coexist for a certain value of $1/\nu$, a 2D branch is
created, whose amplitude is the sum of the 1D amplitudes.}

\includegraphics[%
  bb=54bp 54bp 706bp 628bp,
  scale=0.6]{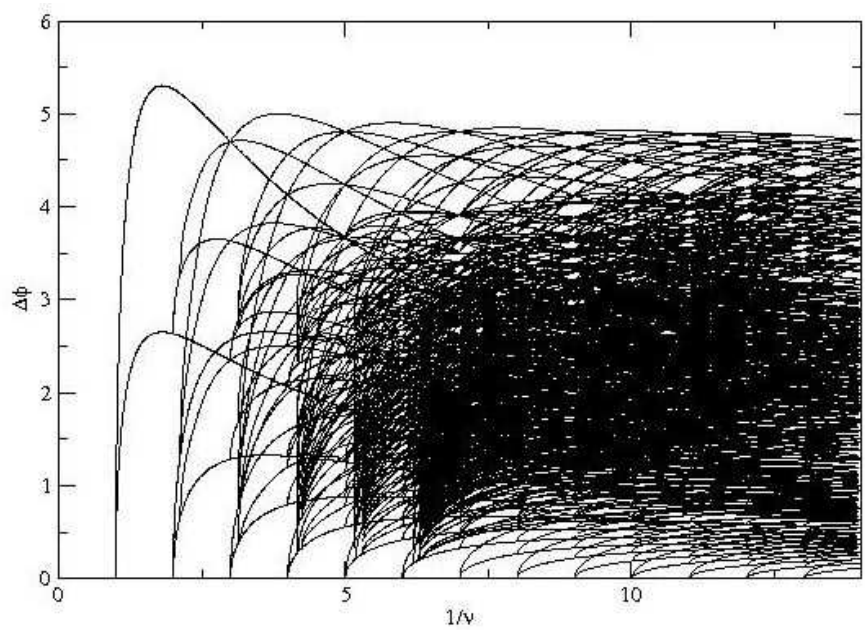}
\end{figure}

The author would like to insist here on different points. First, it
is only possible to obtain such an incredible number of stationary
solutions because of two properties of the Sivashinsky equation, the
pole decomposition, which transforms the search of stationary solutions
in one dimension in a 0D problem, then the possibility to add 1D solutions
in order to get 2D rectangular solutions of the Sivashinsky equation.
In the Kuramoto-Sivashinsky equation case (a non linear equation with
a different growth rate but the same non linear term) the pole decomposition
is not available, but nevertheless a lot of 1D stationary solutions
have been obtained \cite{greenekim}. The Kuramoto-Sivashinsky equation
shares with the Sivashinsky equation the possibility to create 2D
solutions by adding two 1D solutions, so actually in this case we
have also a very large number of branches. These rectangular solutions
are not as physically relevant in the Kuramoto-Sivashinsky equation
case. Contrary to the Sivashinsky equation, where stable stationary
solutions are basically as large as possible and are thus rectangular
in a rectangular domain, it seems likely that in the Kuramoto-Sivashinsky
case, the most interesting solutions would have an hexagonal symmetry
(hexagonal cells are also observed for the Sivashinsky equation with
stabilizing gravity \cite{denetnonlinear}). Stationary solutions
of the Sivashinsky equation with hexagonal symmetry should exist too,
and the author conjectures that the order of magnitude of the number
of solutions with hexagonal symmetry should be approximately the same
as those with rectangular symmetry. Apparently there is no trivial
way to construct hexagonal solutions, so unfortunately, until some
progress is made, obtaining the hexagonal equivalent of Figure \ref{fig:all_2d}
is almost impossible. We have here an example emphasizing the fact
that as the smoothing effect (viscosity, curvature, surface tension
...) decreases, we are not able to generate correctly all the simple
solutions of a given set of partial differential equations (Sivashinsky
and Kuramoto-Sivashinsky equations, Navier Stokes ...) even with the
aid of computers.

\section{evolution with noise \label{sec:evolution-with-noise}}

In the previous section we have shown numerically that the sums of
linearly stable 1D bicoalescent solutions lead to linearly stable
2D solutions. However, even a linearly stable solution could have
a very small basin of attraction. So in this section, we study the
effect of noise on the solutions of the Sivashinsky equation in a
square domain, with Neumann boundary conditions. The important solutions
will be the solutions that are reasonably resistant to the applied
noise.

This noise used here is simply an additive noise, added to the right-hand
side of Equation (\ref{eq:sivashinsky2d}). We choose the simplest
possible noise, a white noise (in space and time), which is gaussian,
has deviation one and is multiplied by an amplitude $a$. But contrary
to Figure \ref{fig:ampli_3,2_4,1}, this noise will be applied at
each time step. We use in all the simulations presented the same parameter
$1/\nu=10$, the stationary solutions corresponding to this parameter
have been presented in the previous section. We recall that in I,
for the one dimensional version of the Sivashinsky equation with moderate
noise, the evolution was analysed in terms of jumps between the available
bicoalescent stationary solutions. We would like to show here that
in 2D, the sums of bicoalescent solutions also play an important role
in the dynamics.

\begin{figure}

\caption{\label{fig:ampli_noise0.01}Amplitude vs time for $1/\nu=10$ and
$a=0.01$ (low noise amplitude). Deexcitation from the $(4,1)\oplus(4,1)$
solution toward the $(3,2)\oplus(3,2)$ solution. This diagram suggests
that the solution is first close to the $(3,2)\oplus(4,1)$ solution,
then from the $(3,2)\oplus(3,2)$ solution, i.e. that the solution
with the lowest amplitude is the most noise resistant.}

\includegraphics[%
  bb=42bp 54bp 789bp 628bp,
  scale=0.6]{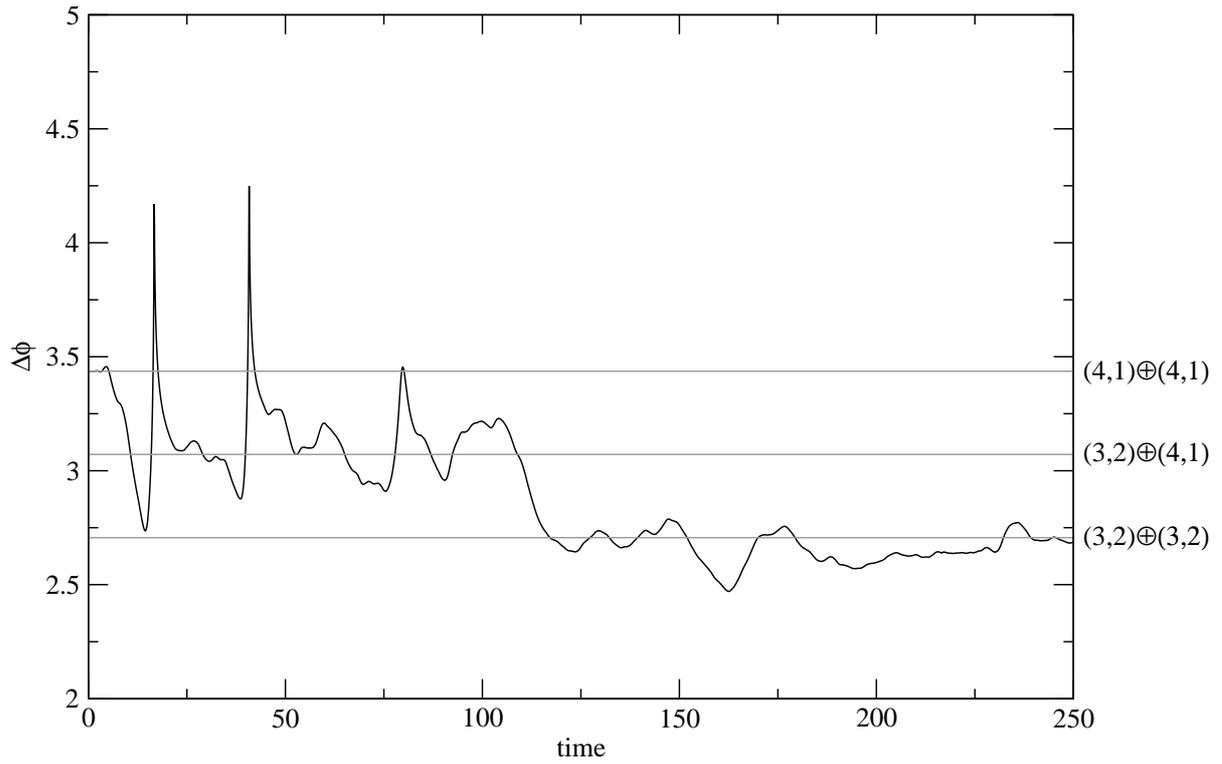}
\end{figure}

In Figure \ref{fig:ampli_noise0.01}, starting from an initial condition
which is the $\left(4,1\right)\oplus(4,1)$ stationary solution, is
plotted the amplitude of the solution versus time, for a noise amplitude
$a=0.01$, with also straight lines corresponding to the amplitudes
of the lowest amplitude stationary solutions, i.e. those of Figure
\ref{fig:3_others}. The stationary solutions with higher amplitudes
(those of Figure \ref{fig:3_5,0}) apparently are too sensitive to
noise to play any role in the dynamics. It is seen in Figure \ref{fig:ampli_noise0.01}
that because of the noise, the solution departs quickly from the $\left(4,1\right)\oplus(4,1)$
solution, and that it seems that, during the time evolution, the solution
is close (apart from some violent peaks in the amplitude) to the $\left(3,2\right)\oplus(4,1)$
solution for some time, then finally the amplitude decreases again
to be near that of the $\left(3,2\right)\oplus(3,2)$ solution. 

\begin{figure}

\caption{\label{fig:norm_noise0.01}Distance to the main stationary solutions
vs time for $1/\nu=10$ and $a=0.01$. A distance is a norm of the
difference between the solution at a given time and the stationary
solution. This diagram makes it possible to verify if a solution at
a given time is indeed close to a stationary solution.}

\includegraphics[%
  bb=45bp 54bp 718bp 628bp,
  scale=0.6]{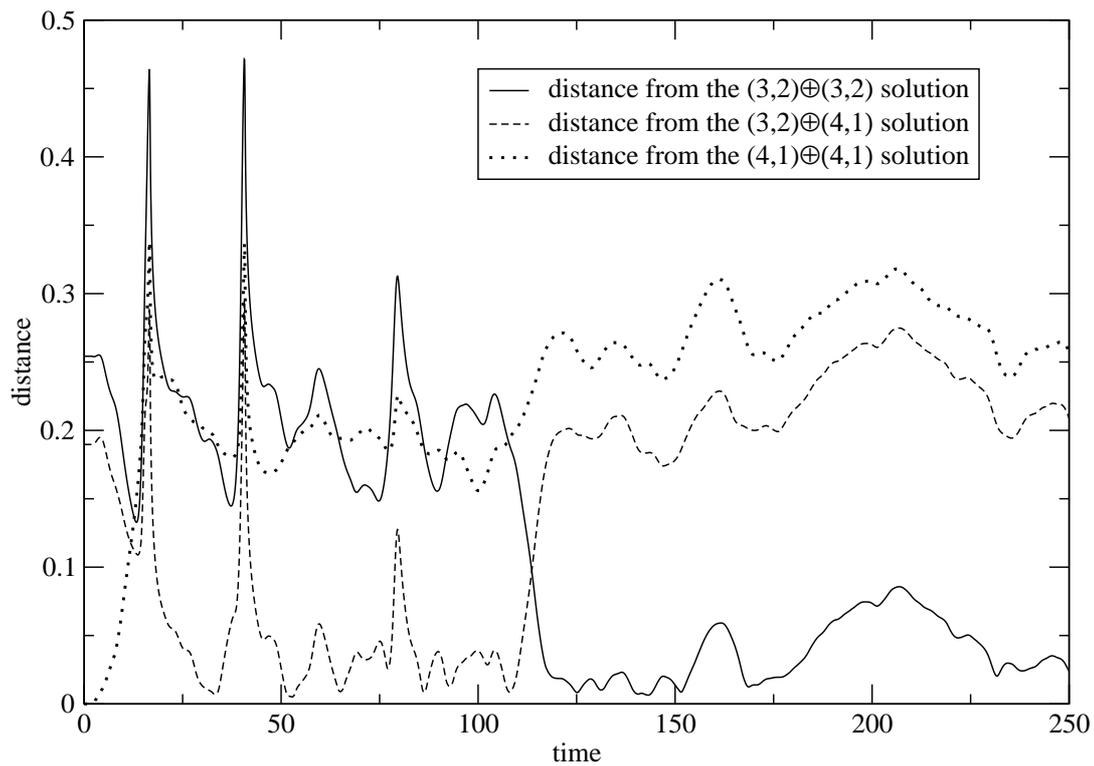}
\end{figure}

In order to prove that the solution is indeed close to the previously
mentioned solutions, because after all very different solutions could
have similar amplitudes, we plot in Figure \ref{fig:norm_noise0.01}
for the same simulation, what we have called the distance between
the solution at a given time and the sums of bicoalescent solutions,
which is simply the L1 norm of the difference between both solutions.
The spatial mean value of all solutions is adjusted here to have the
same value. Normally, it is necessary to measure the distance between
the solution and all symmetries of a given sum of bicoalescent solutions
(i.e. you can interchange the poles at $0$ and $\pi$ in the $x$
and $y$ directions) but for the low amplitude $a=0.01$, it has not
been necessary, and we plot only the distance from the relevant solutions. 

As we start from the $\left(4,1\right)\oplus(4,1)$ solution, the
distance to this solution is zero initially, and we can see that,
although the amplitude seems to indicate that at some time, one is
again close to this solution, this is not the case. On the contrary,
the solution returns regularly close to the $\left(3,2\right)\oplus(4,1)$
solution for times lower than $110$, then there is a transition toward
something close to the $\left(3,2\right)\oplus(3,2)$ solution, the
solution departs slightly from this last solution for some time, possibly
toward a linearly unstable stationary solution, and returns toward
it at the end of the simulation. As Figure \ref{fig:stable_2d} remotely
looks like energy levels in atomic physics, one could be tempted to
interpret the evolution of the two previous figures with a small noise
(apparently in 2D the solution is less sensitive to a given amplitude
of the white noise compared to 1D simulations) as a sort of deexcitation
from the high amplitude level $\left(4,1\right)\oplus(4,1)$ toward
first $\left(3,2\right)\oplus(4,1)$, then toward the fundamental
level $\left(3,2\right)\oplus(3,2)$. Indeed, between the sums of
bicoalescent solutions, if all are linearly stable, the solutions
with the lower amplitude seems to be more resistant to the action
of noise. 

\begin{figure}

\caption{\label{fig:ampli_noise0.1}Amplitude vs time for $1/\nu=10$ and
$a=0.1$ (moderate noise amplitude). This figure suggests that the
solution is often close to the $(3,2)\oplus(3,2)$ solution. It will
be shown in the following figures that it is close to the $(3,2)\oplus(4,1)$
solution only for times around $50$.}

\includegraphics[%
  bb=42bp 54bp 789bp 628bp,
  scale=0.6]{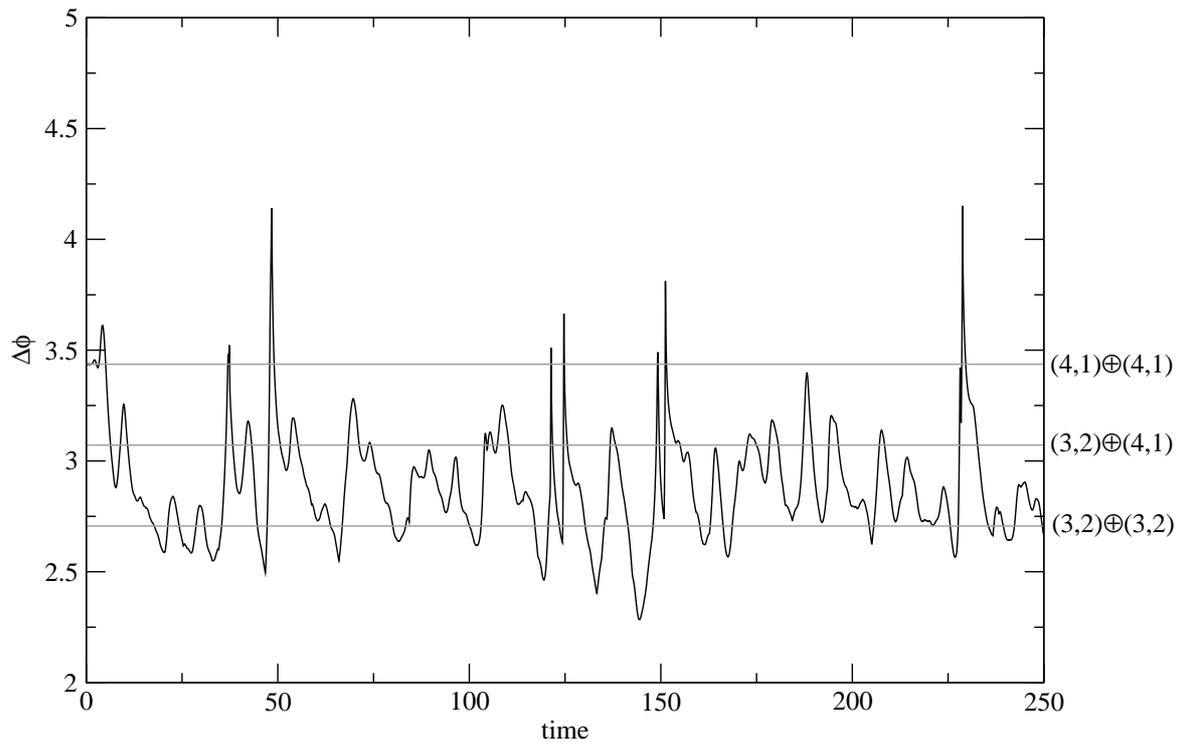}
\end{figure}

To better understand the effect of noise, we present now a simulation
with a larger noise amplitude $a=0.1$ , ten times larger than the
previous case (we recall that this noise amplitude should be compared
to the laminar flame velocity, which is normalized to $1$ in this
paper). In Figure \ref{fig:ampli_noise0.1} is plotted the amplitude
versus time, with as before straight lines with the amplitude of the
important sums of bicoalescent solutions. The initial condition is
also the $\left(4,1\right)\oplus(4,1)$ solution. Apparently this
last solution is too sensitive to noise to play a meaningful role
in the dynamics, although it happens that some peaks in the amplitude
could involve solutions not too far from this initial solution. As
the distance to this solution is never really small, even in the peaks,
we shall not comment further on this solution. On the other hand,
it seems that a lot of time is spent with an amplitude close to that
of the $\left(3,2\right)\oplus(3,2)$ solution (which we have called
previously the fundamental level), and perhaps some time with an amplitude
close to the $\left(3,2\right)\oplus(4,1)$ solution (the first excited
level). 

\begin{figure}

\caption{\label{fig:normaa_noise0.1}Distance to the different symmetries
of the $(3,2)\oplus(3,2)$ solution vs time for $1/\nu=10$ and $a=0.1$.
The noise is sufficiently large to induce transitions between the
different symmetries of this fundamental level.}

\includegraphics[%
  bb=45bp 54bp 718bp 628bp,
  scale=0.6]{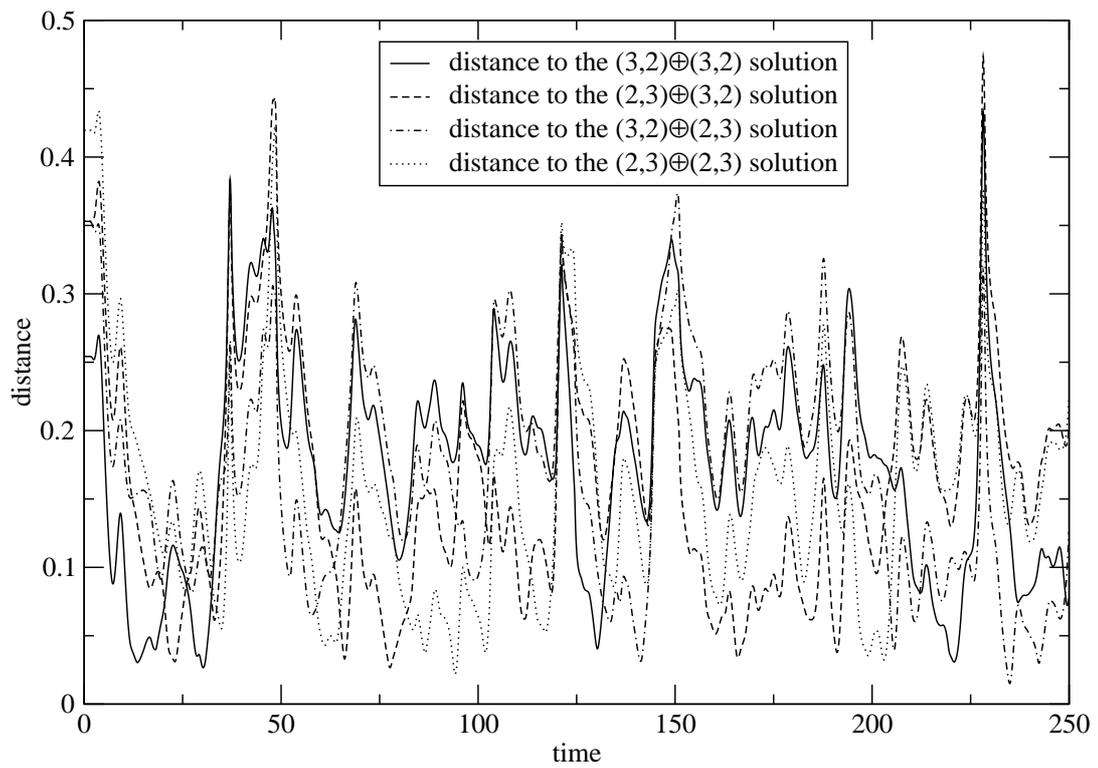}
\end{figure}

In order to see what is really occurring, we now turn to figures of
the distance (defined above) to these two solutions versus time (for
the same simulation of Figure \ref{fig:ampli_noise0.1}). However,
for a higher amplitude, we have to include the four different symmetries
of these solutions in the analysis (i.e. for instance $\left(3,2\right)\oplus(4,1)$
$\left(3,2\right)\oplus(1,4)$ $\left(2,3\right)\oplus(4,1)$ $\left(2,3\right)\oplus(1,4)$
) . In Figure \ref{fig:normaa_noise0.1} is shown the distance to
the four symmetries of the $\left(3,2\right)\oplus(3,2)$ solution.
The distance to one of the four symmetries is indeed often small (but
not very small for this value of the noise) during the time evolution.
Then because of the noise, perturbations are created that lead the
amplitude to increase as the perturbation is convected toward one
of the cusps, and the solution often comes back toward another symmetry
of the fundamental level.

\begin{figure}

\caption{\label{fig:normab_noise0.1}Distance to the different symmetries
of the $(3,2)\oplus(4,1)$ solution (first excited level) vs time
for $1/\nu=10$ and $a=0.1$. The solution is only close to one symmetry
of the $(3,2)\oplus(4,1)$ solution for times around $50$ (after
apparently a transition from a $(3,2)\oplus(3,2)$ solution).}

\includegraphics[%
  bb=45bp 54bp 718bp 628bp,
  scale=0.6]{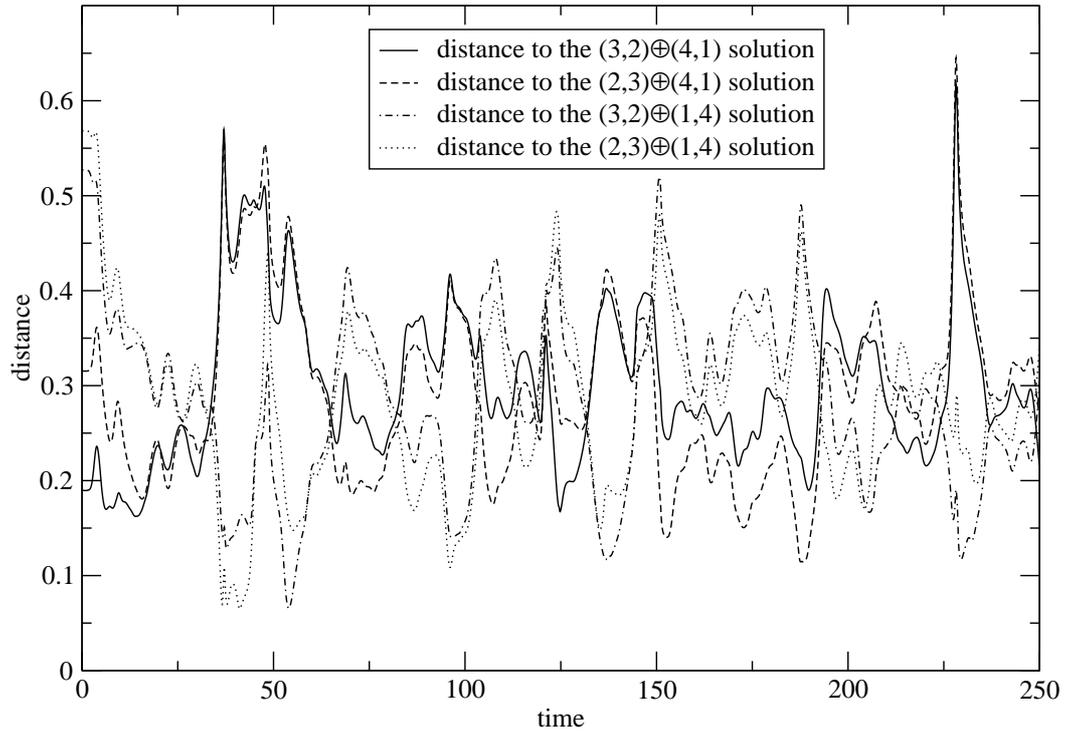}
\end{figure}

In Figure \ref{fig:normab_noise0.1} is shown the distance to the
four symmetries of the $\left(3,2\right)\oplus(4,1)$ solution (the
first excited level) (always for the same simulation). It is seen
that the solution is only reasonably close to this type of solution
at times close to $50$. At other times, minima of the distance are
not very small and the solution is often closer to the $\left(3,2\right)\oplus(3,2)$
solution. In I, we have presented the evolution of the 1D Sivashinsky
equation with a moderate additive noise as a series of jumps between
bicoalescent solutions. In 2D the situation is relatively similar,
with the sums of bicoalescent solutions playing the same role. However,
the noise amplitude necessary to cause jumps seems much higher in
2D, and practically speaking during the previous simulation, only
the fundamental (the solution with the lowest amplitude) and first
excited levels were obtained. It should also be noted that the degenerescence
(the four possible symmetries) of the fundamental level is probably
important in the evolution (for instance for $1/\nu=12$ the fundamental
level would be $\left(3,3\right)\oplus(3,3)$, which does not lead
to other solutions by symmetry, so that it should be less likely to
obtain the fundamental level in this case).

\begin{figure}

\caption{\label{fig:sauve120.1_noise0.1}Solution at time $120.1$ for $1/\nu=10$
and $a=0.1$, plotted as a grayscale figure (white: minimum of $\phi$,
black: maximum of $\phi$). The solution is plotted in the interval
$[0,2\pi]*[0,2\pi]$. Presence of an oblique perturbation which has
grown on a sum of bicoalescent solutions. This perturbation will reach
one corner in the figure, be damped, and the solution will again be
close to a sum of bicoalescent solutions.}

\includegraphics[%
  bb=45bp 54bp 718bp 628bp,
  scale=0.6]{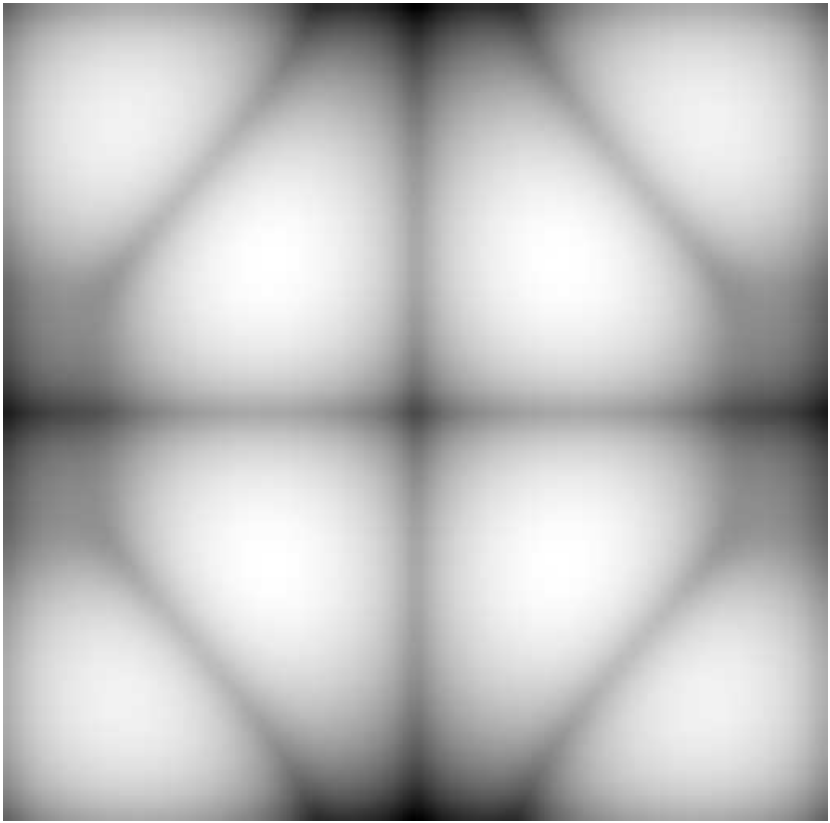}
\end{figure}

Before closing this section, let us insist on the fact that, if the
solution regularly returns toward sums of bicoalescent solutions,
the fronts we obtain are not sums for each time. Figure \ref{fig:sauve120.1_noise0.1},
where is plotted a front of the previous simulation for time $120.1$
(just before a peak of the amplitude in Figure \ref{fig:ampli_noise0.1})
, should be a clear example of this property. In this Figure, the
whole domain $[0,2\pi]*[0,2\pi]$ is plotted as before, but this time
as a grayscale figure, white corresponding to the minimum of $\phi$,
black to the maximum. Essentially, an oblique perturbation has grown
on a front that was previously a sum. This oblique perturbation moves
toward each corner of Figure \ref{fig:sauve120.1_noise0.1}, and the
amplitude peak corresponds to the moment where the perturbation reaches
the corner. Then the solution is attracted again toward a sum of bicoalescent
solutions.

To summarize this section on the effect of noise, the fact that all
the sums of bicoalescent solutions with the optimal number of poles
are linearly stable does not prove that they can be practically observed.
On the contrary, the solutions with the larger amplitude have a basin
of attraction so small that they can almost never be seen. We have
introduced an analogy with atomic physics by calling the bicoalescent
solution with the lowest amplitude the fundamental level, other solutions
the excited levels. In the examples shown, only the fundamental and
first excited levels (and their symmetries) were obtained during the
time evolution of the Sivashinsky equation excited by an additive
noise. We recall that in I, it was shown in 1D that the evolution
with noise was completely different with periodic boundary conditions,
where only the largest amplitude monocoalescent solution was linearly
stable (even if extremely sensitive to noise). In this case, the solution
regularly returns close to the highest amplitude solution. With Neumann
boundary conditions, this is just the opposite, the solution prefers
to be close to the lowest amplitude, almost symmetric, sum of bicoalescent
solutions.

\section{conclusion \label{sec:conclusion}}

In this paper, we have used the possibility to create two dimensional
rectangular stationary solutions from the addition of two 1D stationary
solutions in order to generate a huge number of stationary solutions
of the Sivashinsky equation. With Neumann boundary conditions, the
addition of two stable 1D bicoalescent solutions leads to stable 2D
solutions, which also play a role in the dynamics when an additive
noise is added to the equation. However, with noise, only the sums
of bicoalescent solutions with the lowest amplitude (which are less
sensitive to noise) have a reasonable chance to be observed. More
precisely, jumps between different symmetries of the lowest amplitude
sum, or between the two sums with the lower amplitude, are obtained
in the simulations. Although we have used a white noise in this paper,
experiments, submitted to a residual turbulence, should behave in
a similar way. In order to have a large enough Froude number for gravity
effects to be negligible, flames with a sufficiently large laminar
flame velocity would have to be chosen.

\bibliographystyle{/usr/share/texmf/bibtex/bst/revtex4/apsrev}
\bibliography{./turb2d.reflib}

\begin{figure}

\caption{\label{fig:3_5,0} Perspective view of the (from top to bottom) $\left(5,0\right)\oplus(5,0)$,
$\left(5,0\right)\oplus(4,1)$, $\left(5,0\right)\oplus(3,2)$ stationary
solutions for $1/\nu=10$. The solution is plotted in the interval
$[0,2\pi]*[0,2\pi]$ because it is easier to visualize. Actually,
Neumann boundary conditions are satisfied in $[0,\pi]*[0,\pi]$ (one
fourth of the domain shown).}

\includegraphics[%
  bb=41bp 54bp 718bp 722bp,
  scale=0.7]{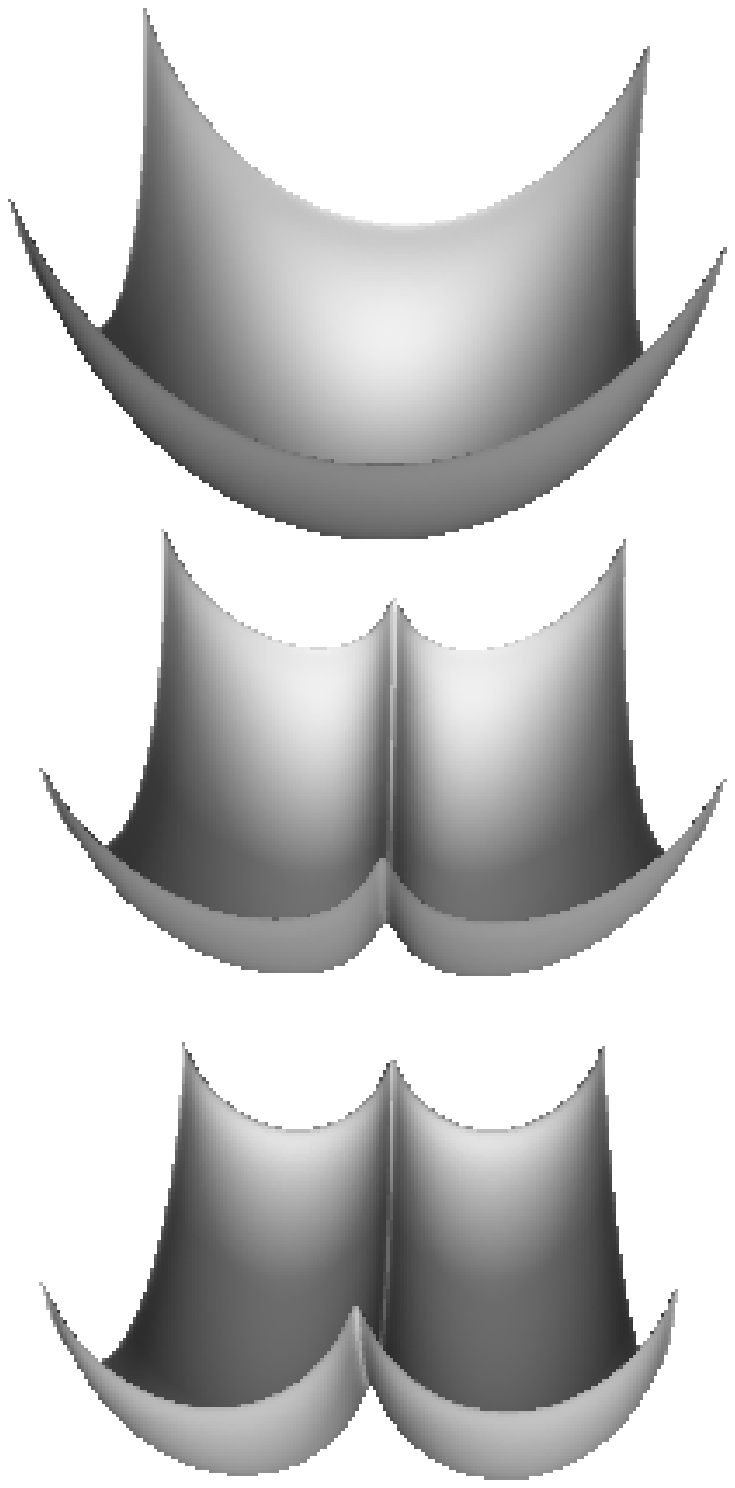}
\end{figure}

\begin{figure}[t]

\caption{\label{fig:ampli_3,2_4,1}Amplitude vs time for $1/\nu=10$ , starting
from a $(4,1)\oplus(3,2)$ solution. A gaussian white noise (amplitude
$a=0.001$ ) is imposed on this solution when time is smaller than
$0.5$. The solution returns exponentially toward the initial solution.}

\includegraphics[%
  bb=42bp 54bp 706bp 628bp,
  scale=0.6]{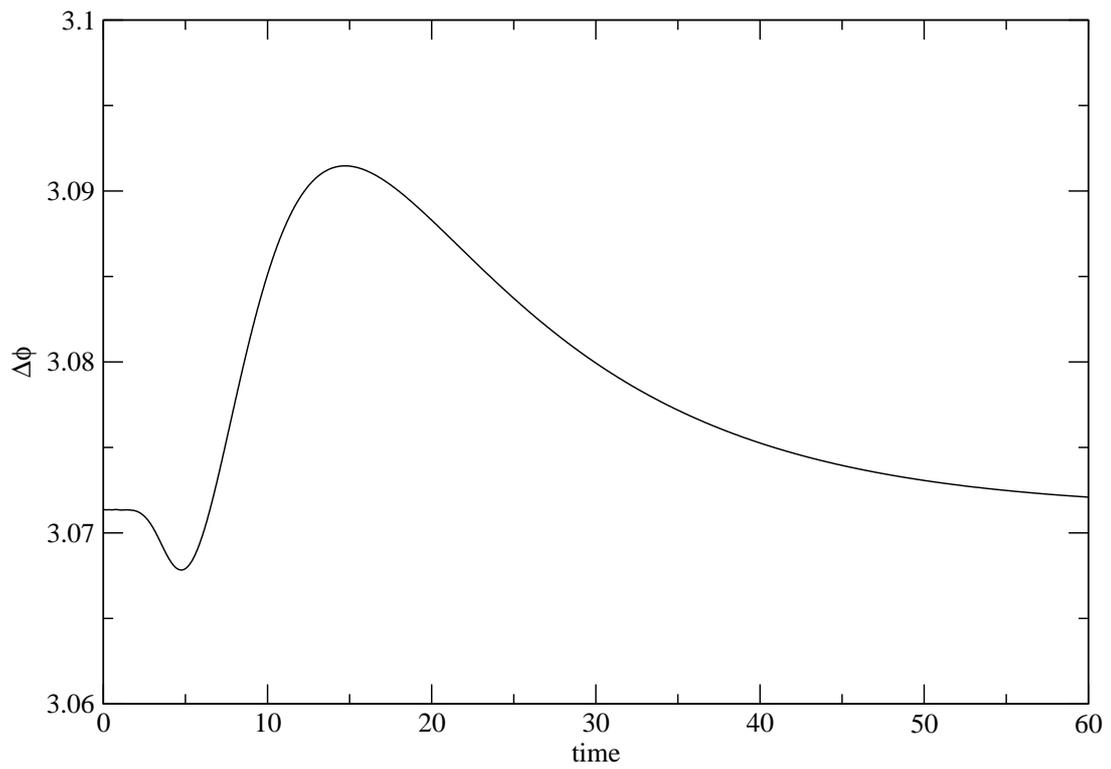}
\end{figure}

\end{document}